\documentclass{PoS}

\usepackage{booktabs}
\RequirePackage{lineno} 
\newcommand{\sNN}{$\sqrt {{s_{\rm NN}}}$}

\title{Off-diagonal cumulants of net-charge, net-proton and net-kaon multiplicity distributions in Au+Au collisions at $\sqrt{s_{NN}}$ = 7.7-200 GeV from STAR}

\ShortTitle{Off-diagonal cumulants}

\author{\speaker{Arghya Chatterjee} {(for the STAR collaboration)}\\
        Variable Energy Cyclotron Centre, HBNI, Kolkata-700064, India\\
        E-mail: \email{arghya@vecc.gov.in}}

\abstract{
Fluctuations of conserved quantities such as net-baryon, net-charge, and net-strangeness numbers have generated considerable interest in the study of the thermodynamic properties of the hot and dense QCD matter. Theoretical calculations suggest that the off-diagonal cumulants of conserved charges along with the diagonal cumulants can help better constrain the freeze-out parameters and, therefore, help to map the QCD phase diagram. In this proceeding, we briefly outline the recent STAR measurements~\cite{Adam:2019xmk} on the second-order off-diagonal cumulants of net-charge, net-proton, and net-kaon multiplicity distributions in Au+Au collisions from the RHIC BES-I program in the energy range of $\sqrt{s_{NN}}$ = 7.7-200 GeV. 
The measured cumulant ratios are compared to the predictions from both thermal (HRG) and non-thermal (UrQMD) models.
}

\FullConference{Corfu Summer Institute 2018 "School and Workshops on Elementary Particle Physics and Gravity"\\
		(CORFU2018)\\
		31 August - 28 September, 2018\\
		Corfu, Greece}

\begin{document}

\section{Introduction}

The major goal of relativistic heavy-ion collision experiment is to study the formation of a new form matter, called the Quark-Gluon Plasma (QGP). Over the last two decades, a number of possible evidences for the QGP phase has been established experimentally. Currently a large community of physicists are exploring the QCD phase structure and trying to find a possible signature of the QCD critical point. One of the most community used method to explore phase diagram is to study the event-by-event fluctuations of conserved charges, such as net electric charge (Q), net baryon number (B) and net strangeness number (S) in the heavy-ion collisions over a wide range of energy~\cite{Stephanov:2008qz, Jeon:2003gk}. It is proposed that the higher-order cumulants of the net-multiplicity distributions are related to the higher order thermodynamic susceptibilities of corresponding conserved charges in QCD and expected to diverge near the critical point~\cite{Stephanov:2008qz, Luo:2015doi}. Over the past few years, STAR and PHENIX experiments at RHIC published results on the diagonal cumulants ($c_{\alpha}^{n}$) of net-electric charge~\cite{Adamczyk:2014fia, Adare:2015aqk}, net-proton ($p$, an experimental proxy for net-baryon)~\cite{Aggarwal:2010wy, Adamczyk:2013dal} and net-kaon ($k$, an experimental proxy for net-strangeness)~\cite{Adamczyk:2017wsl} multiplicity distributions. Similarly, off-diagonal cumulants ($c_{\alpha,\beta}^{m,n}$) of Q, p and k are related to the mixed susceptibilities that carry the correlation between different conserved charges~\cite{Adam:2019xmk}. Lattice QCD and hadron resonance gas model (HRG) calculations show that normalized baryon-strange correlations, that can be expressed as off-diagonal to diagonal cumulant ($C_{BS} = c_{B,S}^{1,1}/c_{S}^{2}$), are expected to be sensitive to the onset of deconfinement~\cite{Koch:2005vg}. Another importance of studying off-diagonal cumulants is that it can also be used to constrain the freeze-out parameters in the QCD phase diagram. Different theoretical calculations demonstrate that the $2^{nd}$-order off-diagonal cumulants show a significant sensitivity to the difference between HRG and lattice calculations~\cite{Bazavov:2012jq, Karsch:2017zzw}.  

In this report, we present the measurements of $2^{nd}$-order diagonal and off-diagonal cumulants of net charge, net proton and net kaon multiplicity distributions in Au+Au collisions ranging in center of mass energy from \sNN~= 7.7 to 200 GeV, with data taken during the first phase of RHIC Beam Energy Scan (BES-I). 
%Observables and notations are defined in section~\ref{oberservables}. We discuss the analysis details in section~\ref{analysis}. The results are discuss in section~\ref{results} and summarize in section~\ref{summary}.  

\section{Observables}\label{oberservables}

The susceptibilities of the conserved charges of a system in thermal and chemical equilibrium (for a grand-canonical ensemble) can be computed from the partial derivatives of the dimensionless pressure with respect to the chemical potentials: 

\begin{eqnarray}
\chi^{m,n,l}_{B,Q,S} = \frac{\partial^{m+n+l} ( P/T^4 )}{\partial^m (\mu_B/T) \partial^n (\mu_Q/T) \partial^l(\mu_S/T)},
\label{eq.sus}
\end{eqnarray}
where $V$ and $T$ are the system pressure and temperature, respectively, and $m, n, l = 1,2,3,...n$ are the order of derivative. $\mu_{Q}$, $\mu_{B}$ and $\mu_{S}$ are the electric charge, baryon and strangeness chemical potentials, respectively. The $P$ is obtained from the logarithm of the QCD partition function:

\begin{eqnarray}
P = \frac{T}{V} \ln [ Z(V,T,\mu_{B},\mu_{Q},\mu_{S})].
\label{eq.pressure}
\end{eqnarray}

These susceptibilities can be related to the cumulants ($c$) of the event-by-event distribution of the associated conserved charges by~\cite{Jeon:2003gk, Majumder:2006nq, Chatterjee:2016mve}: 

\begin{eqnarray}
\chi^{m,n,l}_{B,Q,S} = \frac{1}{VT^{3}} c^{m,n,l}_{B,Q,S}.
\label{eq.cumulants}
\end{eqnarray}
 
 Due to the limitation in detecting all baryons and strange hadrons experimentally, net proton ($p$) and net kaon ($k$) are considered as proxies for the net baryon and net strangeness, respectively. 
 In this report, we present the measurement of  second-order ($m+n+l = 2$) diagonal and off-diagonal cumulants of net charge, net proton and net kaon multiplicity distributions, can be expressed as: 
 
 \begin{eqnarray}
  c_{\alpha}^{2} = \sigma_{\alpha}^{2} = \langle (\delta N_{\alpha})^{2} \rangle~~~~~\textit{and}~~~~c_{\alpha,\beta}^{1,1} = \sigma_{\alpha, \beta}^{1,1} = \langle (\delta N_{\alpha})(\delta N_{\beta}) \rangle,
 \end{eqnarray}
where $\alpha$ and $\beta$ can be $Q$, $p$ or $k$, and $\delta N_{\alpha} = (N_{\alpha^{+}}-N_{\alpha^{-}}) - \langle~(N_{\alpha^{+}}-N_{\alpha^{-}})~\rangle$. Finally, we construct the off-diagonal to diagonal cumulant ratios ($C_{p,k} = c^{1,1}_{p,k} / c^{2}_{k}$, $C_{Q,p} = c^{1,1}_{Q,p} / c^{2}_{p}$ and $C_{Q,k} = c^{1,1}_{Q,k} / c^{2}_{k}$) motivated by Ref.~\cite{Koch:2005vg}, which also cancel the volume dependence.

\section{Experimental details}\label{analysis}

Second-order cumulants of net charge, net proton and net kaon multiplicity distributions for Au+Au collisions at \sNN~= 7.7, 11.5, 14.5, 19.6, 27, 39, 62.4 and 200 GeV have been studied in the STAR experiment. 
%The number of minimum-bias events analyzed at each collision energy is listed in Table~\ref{table:datasets}. 
The minimum-bias (MB) events are analyzed with the requirement that the position of the primary vertex along $z$-direction ($V_{z}$) was reconstructed within $\pm30$ cm of the center of the STAR detector and within 2 cm on the transverse plane of the beam axis. 
The number of events analyzed at each energy after applying all event selections criteria is listed in Table~\ref{table:datasets} . 
\begin{table}[htp!]
	\centering
	\small\addtolength{\tabcolsep}{1pt}
	\begin{tabular}{ c c c}
		\toprule
		\sNN~(GeV) & Year & Events ($\times10^{6}$) \\
		\midrule
		7.7 & 2010 & 1.5 \\
		11.5 & 2010 & 2.5\\
		14.5 & 2014 & 12.7\\
		19.6 & 2011 & 15.6\\
		27 & 2011 & 25.2\\
		39 & 2010 & 62.3\\
		62.4 & 2010 & 31\\
		200 & 2011 & 74\\
		\bottomrule
	\end{tabular}
	\caption {Summary of the number of events analyzed.}
	\label{table:datasets}
\end{table}  

All charged tracks used in this analysis are required to be within the pseudorapidity range $|\eta|< 0.5$, and the transverse momentum range $0.4< p_{\rm{T}}<1.6$ GeV/$c$. To reduce the contamination from the secondary charged particles, only primary tracks are selected within a distance of closest approach (DCA) to the primary vertex of less than 1~cm. 
The main detectors used in this analysis are the Time Projection Chamber (TPC) and the Time of Flight (TOF). The particle identification (PID) is done with a common acceptance: $0.4< p_{\rm{T}}<1.6$ GeV/$c$ and $|\eta|< 0.5$. Within this range, the purities of $K^{\pm}$ and $p(\bar{p})$ identification is estimated to be 98-99\%. 
The collision centrality for this analysis is defined using uncorrected charged particle multiplicity measured within a pseudorapidity range of $0.5 < |\eta| < 1.0$ in the TPC detector. This way we exclude the particles from the analysis region to determine the centrality in order to suppress autocorrelation effects~\cite{Luo:2013bmi}. 
We present the results for nine centrality bins, 0-5\% (most central), 10-20\%, ... , 70-80\% (most peripheral) as a function of average number of participating nucleons ($\langle N_{part} \rangle$) estimated using a Monte Carlo Glauber model~\cite{Abelev:2008ab}. 
The cumulants and their ratios were calculated as a function of the reference multiplicity and then averaged over the centrality bins to suppress the volume fluctuations over wide centrality bins~\cite{Sahoo:2012wn, Luo:2017faz}. 
We use embedding Monte Carlo simulation techniques to obtain the efficiencies and an algebra based on binomial detector response to efficiency correction~\cite{Kitazawa:2012at, Bzdak:2012ab, Luo:2014rea}. The statistical uncertainty estimation is based on the numerical error propagation method of multivariate cumulants~\cite{smomentPT}. The systematic uncertainties are estimated by varying different track quality cuts, tracking efficiency and conditions for particle identification.

\section{Results}\label{results}

\begin{figure*}[htp!]
	\vspace{0.5cm} 
	\centering 
	\includegraphics[scale=0.42]{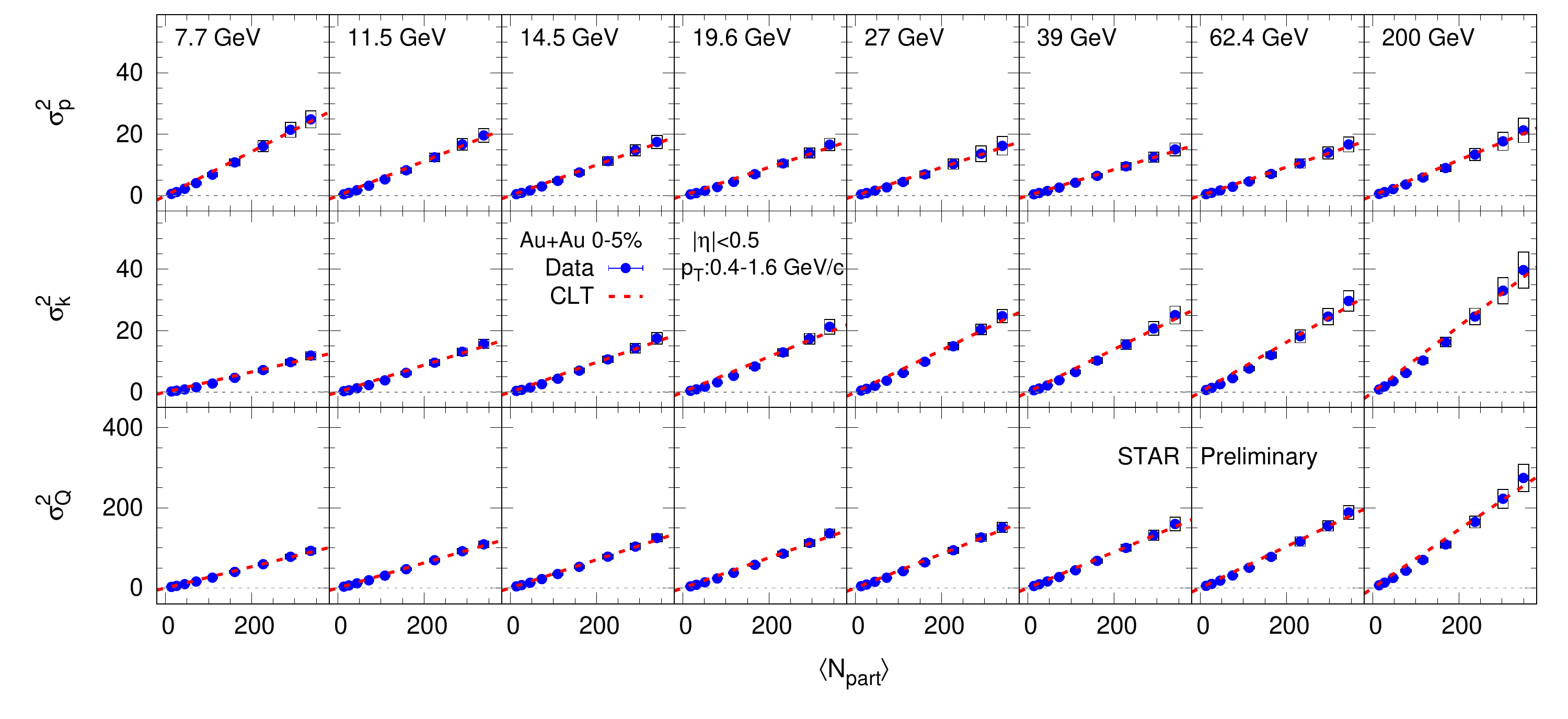}
	\caption{Centrality dependence of the second-order diagonal cumulants (variances) of net proton, net kaon and net charge (top to bottom) multiplicity distributions for Au+Au collisions at \sNN~= 7.7-200 GeV (left to right) within kinematic range $|\eta|<0.5$ and $0.4<p_{T}<1.6$ GeV/$c$. Boxes represent systematic uncertainties. The statistical error bars are within the marker size. The red dashed lines represent a scaling predicted by central limit theorem.}
	\label{variance}
	\vspace{0.5cm} 
\end{figure*}

The centrality dependence of efficiency corrected second-order diagonal cumulants of net proton, net kaon and net charge (top to bottom) for 0-5\% most central Au + Au collisions at \sNN~= 7.7, 11.5, 14.5, 19.6, 27, 39, 62.4 and 200 GeV are shown as a function of $\langle N_{part} \rangle$ in Fig.~\ref{variance} . We find a linearly increasing trend as expected from a scaling, predicted by the central limit theorem. In a given centrality, the width of net-proton distribution decreases as a function of beam energy in the range \sNN~= 7.7-39 GeV and then increases at top RHIC energies~\cite{Adamczyk:2013dal}. This is because of baryon transport that has a strong beam energy dependence. 

\begin{figure*}[htp!]
	\centering 
	\includegraphics[scale=0.42]{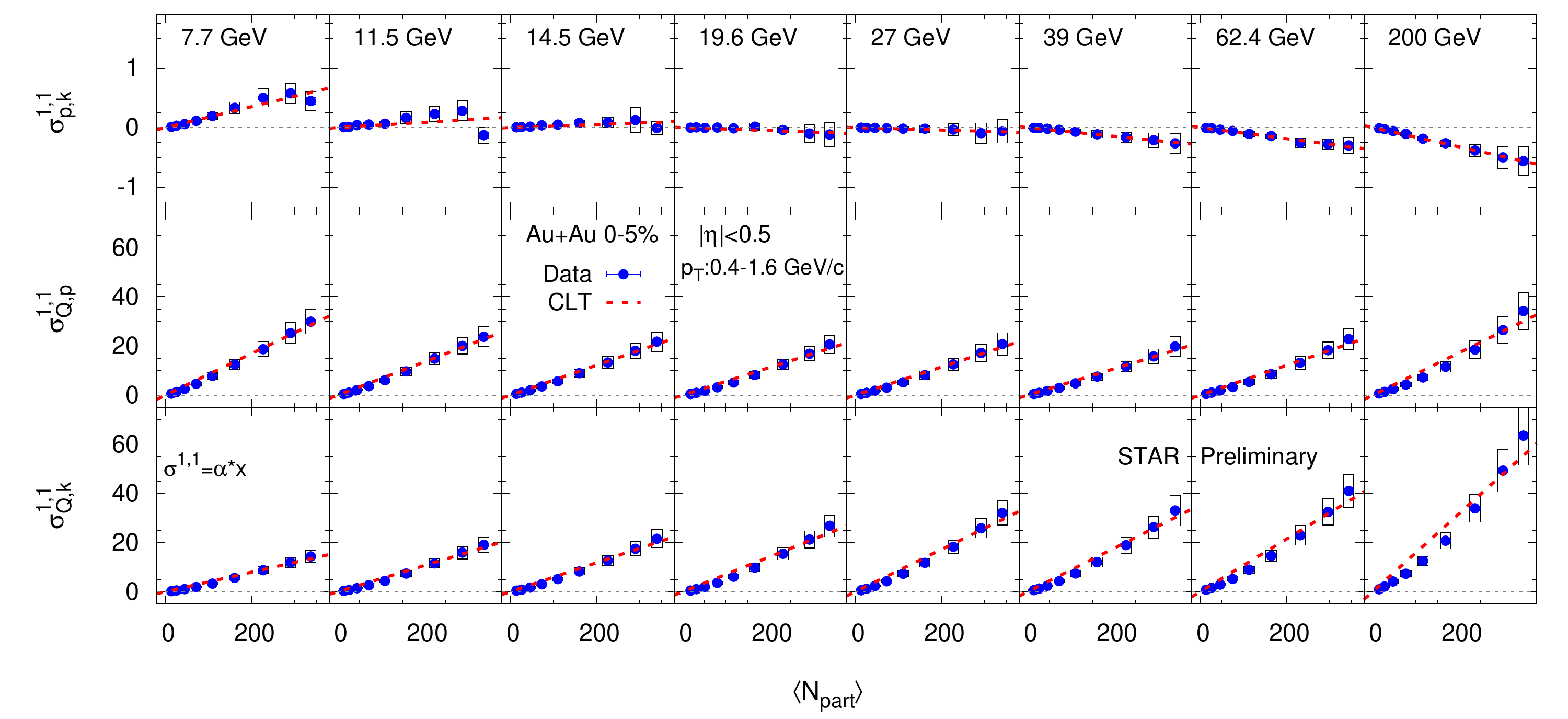}
	\caption{Centrality dependence of the second-order off-diagonal cumulants (covariances) of net-proton, net-charge and net-kaon for Au+Au collisions at \sNN~= 7.7-200 GeV (left to right) within kinematic range $|\eta|<0.5$ and $0.4<p_{T}<1.6$ GeV/$c$. Boxes represent systematic uncertainties. The statistical error bars are within the marker size. The red dashed lines represent a scaling predicted by central limit theorem.}
	\label{covariance}
\end{figure*}

Figure~\ref{covariance} shows the efficiency corrected second-order off-diagonal cumulants of net-charge, net-proton, net-kaon multiplicity distributions for Au+Au collisions at eight colliding energies. The off-diagonal cumulants between net-charge$-$net-kaon ($\sigma^{1,1}_{Q,k}$) and that of net-charge$-$net-proton ($\sigma^{1,1}_{Q,p}$) increase with centrality. On the contrary, there is a growing anti-correlation behaviour observed between net-proton and net-kaon ($\sigma^{1,1}_{p,k}$) with centrality at \sNN~$>$ 19.6 GeV.

\begin{figure}[htp!]
	\centering 
	\includegraphics[scale=0.54]{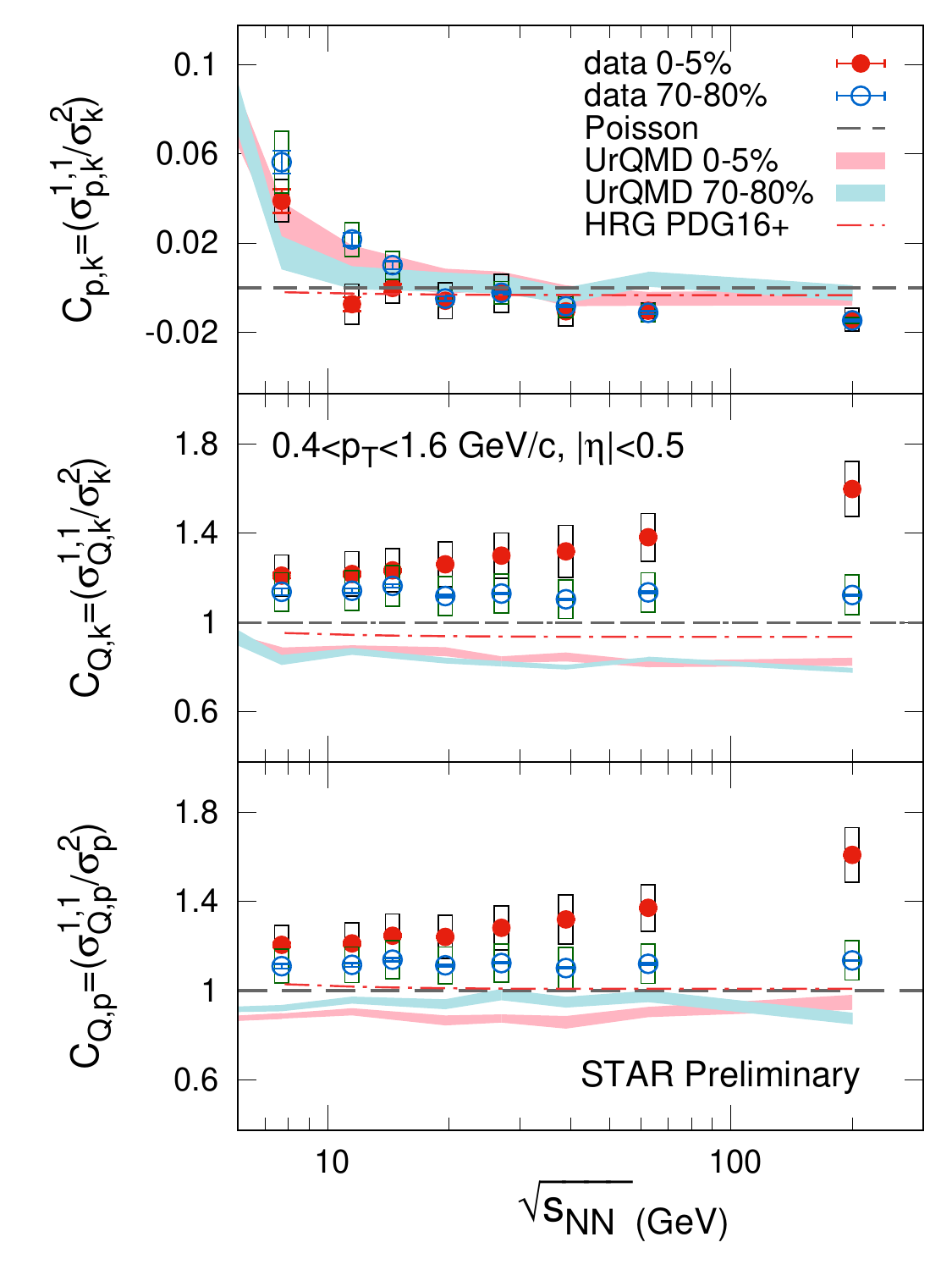}
	\caption{Beam energy dependence of cumulants ratios $C_{p,k}$, $C_{Q,k}$ and $C_{Q,p}$ (top to bottom) for Au+Au collisions at \sNN~= 7.7-200 GeV (left to right). The bands denote the UrQMD data for 0-5\% and 70-80\% central collisions and Poisson baseline is denoted by dotted lines. Error bars and boxes represent statistical and systematic uncertainties, respectively.}
	\label{energy} 
\end{figure}

Figure~\ref{energy} shows the off-diagonal to diagonal cumulant ratios $C_{p,k}=\sigma^{1,1}_{p,k}/\sigma^{2}_{k}$, $C_{Q,k}=\sigma^{1,1}_{Q,k}/\sigma^{2}_{k}$ and $C_{Q,p}=\sigma^{1,1}_{Q,p}/\sigma^{2}_{p}$ (top to bottom) as a function of beam energy for most central (0-5\%) and peripheral (70-80\%) collisions. These cumulant ratios are designed to eliminate the effect of system volume. 
An excess correlation is observed in $C_{Q,p}$ and $C_{Q,k}$ in 0-5\% most central in comparison to the peripheral collisions. The values of $C_{Q,p}$ and $C_{Q,k}$ are observed to increase with beam energy, and this increasing trend cannot be explained by the HRG and UrQMD model calculations. It is observed that the normalized $p$-$k$ correlation ($C_{p,k}$) is positive at the lowest BES energy and negative at higher energies. For 0-5\% top central bins, $C_{p,k}$ changes sign around 19.6 GeV.

\section{Summary}\label{summary}

The second-order diagonal and off-diagonal cumulants of net proton, net kaon, and net charge multiplicity distributions in Au+Au collisions from the RHIC BES-I program in the energy range of $\sqrt{s_{NN}}$ = 7.7-200 GeV are presented. Significant excess correlation is observed in $C_{Q,p}$ and $C_{Q,k}$ in central in comparison with peripheral events. Both HRG and UrQMD model underpredict the data and cannot describe the increasing with beam energy trends of $C_{Q,p}$ and $C_{Q,k}$. The value of $C_{p,k}$ in 0-5\% central collision is found to be negative at \sNN~= 200 GeV and positive at \sNN~= 7.7 GeV. The measurements of the full second-order cumulant matrix elements of net-$p/k/Q$ multiplicity distributions as a function of centrality and beam energy will improve the estimation of freeze-out parameters by theoretical calculations and that help to map the QCD phase diagram. For more details of this analysis we refer the readers to Ref.~\cite{Adam:2019xmk}.

%\begin{thebibliography}{99}
%\bibitem{...}
%....
%\end{thebibliography}

\bibliographystyle{JHEP}
\bibliography{OffDiagonalCPOD}

\end{document}